\documentstyle[sprocl]{article}

 \def \2{\textstyle {1\over 2}}
 \def \3{\textstyle {1\over 3}}

\begin{document}

\title{EFFECTIVE CROSS SECTIONS AND SPATIAL STRUCTURE OF THE HADRONS}
\author{G. CALUCCI, M. DAZZI}

\address{Dipartimento di Fisica teorica, Universit\`a di Trieste, Strada
Costiera 11. I-34014 and I.N.F.N. Sezione di Trieste, Italy } 

\maketitle\abstracts{ Informations about the spatial structure of parton
distribution within the hadron are provided by the ratios between the inclusive
cross sections
for a pair of jets, two pairs of jets, three pairs of jets..., and so on. 
It results, however that these ratios depend not only
on the spatial distribution but also, and even more, on the multiplicity
distribution of the initial partons.}

\section{ Motivations and overall description}
\vskip .5pc
The multiple production of pairs of jets in the high-energy hadron-hadron
collisions provides a way to investigate the spatial, and also non spatial,
structures of the hadrons. A particular instance of this investigation is
presented here, taking into account the inclusive cross sections 
integrated over the momentum spectrum.
\par
When the momentum transfer to the pair of jets $\Delta p$ is large
enough, $i.e.\; 1/\Delta p\,<<\,R_H$, where $R_H$ is the hadron radius a
description of the process in term of the impact parameter $b$ is justified.
This possibility simplifies many calculations and is moreover very useful in
visualizing the processes one wish to study.
\par
In absence of a well established non perturbative QCD, a lot of models may be
proposed, all of them stemming from the original partonic 
description. Using the impact parameter language a list of features 
which are related to the more general aspects of the experimental evidences 
is presented:
\par\noindent
The distribution of the partons in the transverse plane ($b$-distribution).
\par\noindent
The distribution in the fractional longitudinal momenta ($x$-distribution).
\par\noindent
The correlations between transverse and longitudinal variables of the parton: 
$(\vec b,x)$ and among different partons: $(\vec b_1,\vec b_2),\,(x_1,x_2)$.
\par\noindent
The multiplicity distributions of the incoming partons.
\par
Other features like spin, color, flavor distributions involve clearly finer
experimental analyses.
\par
The starting point for the analysis will be, then, the inclusive cross sections
for multiple pair production
$ d^k\sigma/dp_1\dots dp_k $:
integrating these expressions over the relevant kinematical variables we get the 
integrated inclusive cross sections: 

\begin {equation}\sigma_1=<n>\sigma_H,\cdots ,
\sigma_k= <n(n-1)\cdots (n-k+1)>\sigma_H \,,\end
{equation}
where $\sigma_H$ is the  hard contribution to the inelastic cross section.
\par
In order to connect these general definition with the model analysis that will
 be presented, an expression for $\sigma_H$ is given here, in a particularly 
 simple case

\begin {eqnarray} 
\sigma_H^{IJ}&=\int d^2{ \beta} \int
\sum_{n=1}^{\infty}
{1\over n!} \Gamma_I(x_1,{\vec b}_1)\dots \Gamma_I(x_n,{\vec b}_n)
e^{-\int \Gamma_I(x,{\vec b})dx d^2b} \cr\cr
\times & \sum_{m=1}^{\infty}{1\over m!} 
\Gamma_J(x_1',{\vec b}_1'-{\vec \beta})\dots \Gamma_J
(x_m',{\vec b}_m'-{\vec \beta})
 e^{-\int \Gamma_J(x',{\vec b}')dx' d^2b'} \cr\cr
 \times &\Bigl[1-\prod_{i=1}^n\prod_{j=1}^m(1-\hat{\sigma}_{ij})\Bigr]
  dx_1d^2b_1\dots dx_nd^2b_ndx_1'd^2b_1'\dots dx_m'd^2b_m'
   \end {eqnarray}
In this first model the distribution of the partons in the hadron is Poissonian
and completely uncorrelated, different distributions will be considered in
the following. The integration over the fractional momenta
must present a lower bound in order that the parton scattering may involve a
finite momentum transfer, larger than a fixed threshold.
\par	The expression in square parentheses of Eq.(2) represents the 
probability of having at least one semi-hard partonic interaction between 
hadron $I$ and hadron $J$, the expression
 $\hat{\sigma}_{ij}\equiv \hat{\sigma}_{ij}({\vec b}_i-{\vec b}_j;x_i,x_j)$ 
is the probability of having 
a semi-hard interaction of the parton $i$ from hadron $I$ with the parton $j$ 
of the hadron $J$, so it 
depends on $x_i,x_j$, and on the difference of the transverse relative distance
${\vec b}_i- {\vec b}_j'$, according to the considerations made at the beginning
the expression will be taken as local in $\vec b$:
 $\hat\sigma=\bar\sigma_{x,x'} \delta (\vec b-\vec b')$.
The cross section results from the sum over all possible
partonic configurations of the two hadrons followed by the integration on the 
overall hadronic impact parameter $\beta$. 
\par	
One will notice that in Eq.(2) all possible interactions between
partons of hadron $I$ and partons of hadron $J$ are taken into account, so
that also all possible hard elastic rescatterings are included. 
\par
A relevant simplification is obtained by neglecting every rescattering, $i.e.$
by saying that a given parton will interact only once[1]. In that case the
expression in the square parentheses in eq.(2) is simplified to:
\begin{equation}1-\prod_{i=1}^n\prod_{j=1}^m(1-\hat{\sigma}_{ij})\approx
\sum_{i,j}\hat{\sigma}_{ij}-{1\over 2}\sum_{i\neq i' \atop j\neq j'}
\hat{\sigma}_{ij}\hat{\sigma}_{i'j'}+{1\over {3!}}
\sum_{i\neq i'\neq i'' \atop j\neq j'\neq j''}
\hat{\sigma}_{ij}\hat{\sigma}_{i'j'}\hat{\sigma}_{i''j''}-\cdots
\end {equation}
\vfill
\eject
In the same way the inclusive cross section for production of $k$ pairs of jets
with momentum transfer $p_1,\dots,p_k$ is given by

\begin {eqnarray}d^k &\sigma/dp_1\dots dp_k =\int d^2{ \beta} \int
\sum_{n=1}^{\infty}
{1\over n!} {n\choose k}\Gamma_I(x_1,{\vec b}_1)\dots \Gamma_I(x_n,{\vec b}_n)
e^{-\int \Gamma_I(x,{\vec b})dx d^2b} \cr\cr
\times & \sum_{m=1}^{\infty}{1\over m!} {m\choose k}
\Gamma_J(x_1',{\vec b}_1'-{\vec \beta})\dots \Gamma_J
(x_m',{\vec b}_m'-{\vec \beta})
 e^{-\int \Gamma_J(x',{\vec b}')dx' d^2b'} \cr\cr
 \times k!&d\hat\sigma/dp_1\cdots d\hat\sigma/dp_k\,
  dx_{k+1}d^2b_{k+1} \dots dx_nd^2b_n\, dx_{k+1}'d^2b_{k+1}'\dots
  dx_m'd^2b_m'\,,
  \end {eqnarray}
  $i.e.$ $k$ partons from the hadron $I$, $k$ partons from the hadron 
  $J$ are chosen and connected in all the $k!$ ways with the elementary cross
  sections $d\hat\sigma/dp$ the remaining variables are integrated without 
  any constraint.
  A further integration over the kinematical variables $p_1,\dots,p_k$ 
  gives the integrated inclusive cross section, in this particular case:
  \begin {equation} \sigma_k={1\over k!}\int d^2\beta\Bigl[\int 
  \Gamma_I (\vec b,x) \bar\sigma_{x,x'}
  \Gamma_J (\vec b-\vec\beta, x') d^2b dxdx'\Bigr]^k \,.\end {equation}
 The effective cross section is introduced in the usual way[2]:
  \begin{equation} \sigma_{\rm eff}={{\sigma_1^2}\over {2\sigma_2}}
  \end {equation} and the generalizations  for higher integrated inclusive 
  cross sections may be defined in term of dimensionless parameters $\tau_k$
  \begin {equation} \sigma_k={{(\sigma_1)^k}\over {k! (\sigma_{\rm eff})^{k-1}
  \tau_k}}\,. \end{equation}
  In this simplified treatment it is clear that $\sigma_{\rm eff}$ is mainly
  connected with the geometrical properties of the hadron, in fact if 
  $\bar\sigma$ is multiplied by a constant then $\sigma_{\rm eff}$ remains
  unaffected, this property holds also for the parameters $\tau_k$.
  \par
  The relevance of the effective cross section $\sigma_{\rm eff}$ has been
  discussed in another talk[3], here the attention is concentrated on the
  parameters $\tau_k$.
  \vskip .5pc
  \section {Examples}
  \vskip .5pc
  The population of partons, whichever may be its detailed shape, certainly
  increases with decreasing $x$. When the total energy is so
  high that hard scatterings can occur even between low-$x$ partons, 
  these processes are more likely than those involving 
  the few valence quarks. This suggests a further simplification obtained by 
  performing 
  an integration in $x$ of the distribution, assuming
  that the kinematical constraints over the $x$ variables are not very relevant
  precisely because the small-$x$ processes are the dominant ones. The
  expression in eq.(5) is substituted*\footnote 
  {The expression $\bar\Gamma_I$ represents the effect of an integration in $dx$,
  the "bar" will be omitted in the following.}
  with: 
 
   \begin {equation} \sigma_k={1\over k!}\int d^2\beta\Bigl[\int 
  \bar\Gamma_I (\vec b) \bar\sigma
  \bar\Gamma_J (\vec b-\vec\beta) d^2b\Bigr]^k \,.\end {equation}
  
  It is now easy to proceed with the actual calculation choosing some definite
  forms for $\Gamma$. Two choices, easy to treat and different
  enough to allow a first exploration, are:
  
   \begin {equation} \Gamma_G=\rho {1\over {\pi R^2}} \exp [-b^2/R^2]\quad,\quad 
   \Gamma_D=\rho{1\over {\pi R^2}}\theta (R^2-b^2).\end {equation}
  
   In terms of these choices the corresponding values of $\tau_3\,,\,\tau_4$ are
   computed.
   \par 
  The parton population that has been considered till now completely lacks
  correlations among the partons.
  \par
  A simple but efficient way of introducing correlation, that allows also a
  model interpretation is to build up the parton populations in terms of two
  clusters, having their centers spread over the hadron size*\footnote
  {This description has some similarities with the valon model of R.C.Hwa[4];
  however the main attention is directed there to the longitudinal variables,
  here to the transverse variables.}
   To be definite a
  term of this distribution is written as:
  \begin{equation}{1\over {n'!\,n''!}} \int d^2\! B'\,d^2B'' f (B')f (B'')
  \Gamma(\vec b'_1-B')\dots \Gamma(\vec b'_{n'}-B')
  \Gamma(\vec b''_1-B'')\dots \Gamma(\vec b''_{n''}-B'')
  \end {equation}
  with $\int d^2B f (B)=1$.
  In so doing the Poissonian character of the integrated distributions is
  preserved but correlations in the impact parameter are introduced.
  The actual calculation is performed by choosing:
  \begin {equation} \Gamma=\rho {1\over {\pi r^2}} \exp [-(\vec b-\vec B)^2/r^2]
  \quad,\quad f ={1\over {\pi R^2}} \exp [-B^2/R^2]\end {equation}
    One verifies that correlations, in
   form of dependence on $\vec b'-\vec b''$ are introduced by the integration
   over $B$.
   The values of $\tau_3\,,\,\tau_4$ are explicitly computed, they depend on
   the ratio $u=(R/r)^2$.
   \par
   A definite way of departing from the Poisson distribution is to change
   the original weights of the multiplicities, the general term of the
   non correlated distribution:
   \begin {equation} {\cal N}(C_j)
   {C_n\over n!} \Gamma({\vec b}_1)\dots \Gamma({\vec b}_n)
    \end {equation}
    requires some manageable choice of the coefficients $C_j$, in
    particular an explicit form of ${\cal N}(C_j)$ is needed.
    A possible choice is a negative binomial distribution for the initial
    partons*\footnote
    {This kind of distribution was proposed, in a different
    context[5] a long time ago. }
    ; it gives for the coefficients and for the normalization term:
    \begin{equation}C_n=(\nu)_n\equiv \nu(\nu+1)\cdots(\nu+n-1)\quad ,\quad
    {\cal N}(C_j)=\Bigl[1-\int \Gamma(\vec b) d^2b\Bigr]^{\nu} \end{equation}
    
    The Poisson distribution is reached in the limit
    $\rho\to \rho/\nu$ and then $\nu\to\infty$. In this way it is possible 
    to measure how much the results deviate from the
    previous ones when the distribution deviates from the Poissonian form.
    The shape in $\vec b$ of the parton distribution enters in a way which is
    independent of the choice of the coefficients $C_n$, so different choices,
    $e.g.$ the ones of eq. (11), are possible. The calculation of the quantities
    $\tau$ is a bit more laborious than in the Poissonian case, anyhow it can
    be carried out explicitly.
   
   \vskip .5pc
  \section {Numerical results and conclusions}
  \vskip .5pc
   The numerical results for the quantities $\tau$ are presented in Table 1.
   
\begin{table}[h!]
\caption{Values of $\tau$ for different parton populations}
\vspace{0.2cm}
\begin{center}
\large
\begin{tabular}{|c|c|c|c|c|}
\hline
{} &\raisebox{0pt}[13pt][7pt]{$A_1$} &
\raisebox{0pt}[13pt][7pt]{$A_2$} &
\raisebox{0pt}[13pt][7pt]{$B$} &
\raisebox{0pt}[13pt][7pt]{$C$} \\
\hline
\raisebox{0pt}[18pt][7pt]{$\tau_3$} &
\raisebox{0pt}[18pt][7pt]{${3\over 4}$} &
\raisebox{0pt}[18pt][7pt]{0.80} &
\raisebox{0pt}[18pt][7pt]{${3\over 4}\cdot F_3 (u)$} &
\raisebox{0pt}[18pt][7pt]{${3\over 4} \Bigl[{{\nu +1}\over {\nu +2}}\Bigr]^2$}
 \\
\hline
\raisebox{0pt}[18pt][7pt]{$\tau_4$} &
\raisebox{0pt}[18pt][7pt]{${1\over 2}$} &
\raisebox{0pt}[18pt][7pt]{0.56} &
\raisebox{0pt}[18pt][7pt]{${1\over 2}\cdot F_4(u)$} &
\raisebox{0pt}[18pt][7pt]{${1\over 2}
\Bigl[{({\nu +1})^2\over {(\nu +2)(\nu +3)}}\Bigr]^2 $} \\
\hline

\end{tabular}
\end{center}
\end{table}

   \par\noindent
   The column $A_1$ corresponds to an uncorrelated Poissonian distribution and 
   Gaussian shape $\Gamma_G$ in eq.(9).
   \par\noindent
   The column $A_2$ corresponds to an uncorrelated Poissonian distribution and 
   rigid disk shape $\Gamma_D$ in eq.(9).
   \par\noindent
   The column $B$ corresponds to a Poissonian distribution with correlations and 
   Gaussian shape, eqs. (10,11).
   \par\noindent
   The column $C$ correspond to an uncorrelated negative binomial distribution 
   and Gaussian shape $\Gamma_G$, eqs. (12,13).
   \par 
   The functions $F_3,F_4$ which appear in column $B$ are rational functions of
   $u$; both are equal to 1 when $u=0$ and when $u\to\infty$, moreover it
   results $F_3(1)=1.09\,,\,F_4(1)=0.93$, it may be verified in general that
   they do not vary very much; for this reason the more natural case with three
   clusters[4] was not worked out. The square parentheses in column $C$, which 
   are
   always less than 1, may differ strongly from unity for small values of $\nu$,
   $i.e.$ for distributions that differ much from the Poissonian.
   \par
   From this preliminary analysis it results that the higher order integrated
   inclusive cross
   sections feel, obviously, 
   all the characteristics of the parton distribution, but they are mainly
   affected by the multiplicity distribution of the incoming partons and less by
   the spatial shape or by the the spatial correlations of the parton 
   distribution.

\section*{Acknowledgments}
This work is part of a wider investigation performed together with 
D.Treleani.
Discussions with M.A. Braun and R.C. Hwa during ISMD99 are
acknowledged.
\par 

This work has been partially supported by the Italian Ministry of University
 and of Scientific and Technological Research by means of the {\it Fondi per la
 Ricerca scientifica - Universit\`a di Trieste }.

\vfill
\eject  

\section*{References}

\begin{thebibliography}{99}
\bibitem{[1]}
G. Calucci and D. Treleani, {\it Phys. Rev.} {\bf D57} 503 (1998)
\bibitem{[2]}
P.V. Landshoff and J.C. Polkinghorne, {\it Phys. Rev.} {\bf D18} 3344 (1978);
Fujio Takagi {\it Phys. Rev. Lett.} {\bf 43}, 1296 (1979); 
C. Goebel, F. Halzen and D.M. Scott, {\it Phys. Rev.}
{\bf D22}, 2789 (1980);
N. Paver and D. Treleani, {\it Nuovo Cimento} {\bf A70},
215 (1982); B. Humpert, {\it Phys. Lett.} 
{\bf B131}, 461 (1983); M. Mekhfi, {\it Phys. Rev.} {\bf D32}, 2371 (1985),
{\it ibid.} {\bf D32}, 2380 (1985); 
B. Humpert and R. Odorico, {\it Phys. Lett.} {\bf 154B}, 211 
(1985); T. Sjostrand and
M. Van Zijl, {\it Phys. Rev.} {\bf D36}, 2019 (1987);
F. Halzen, P. Hoyer and W.J. Stirling {\it Phys.Lett.} {\bf 188B}, 375 (1987);
M. Mangano, {\it Z. Phys.} {\bf C42}, 331 (1989); 
R.M. Godbole, Sourendu Gupta and J. Lindfors, {\it Z. Phys.} {\bf C47} 69 
(1990).
\bibitem{[3]}G. Calucci and D. Treleani, these proceedings and
G. Calucci and D. Treleani, {\it Phys. Rev.} {\bf D60} 054023-1 (1999)

\bibitem{[4]}R.C. Hwa {\it Phys. Rev.} {\bf D22} 1593 (1980)
\bibitem{[5]}A. Giovannini and L. van Hove {\it Z. Phys.} {\bf C30} 391 (1986).
\end{thebibliography}
\end{document}